# Logistic map and micro-structure of isotropic turbulent flow

**Zheng Ran**

Shanghai Institute of Applied Mathematics and Mechanics,

Shanghai University, Shanghai 200072, China

ONE of the main goals in the development of theory of chaotic dynamical system has been to make progress in understanding of turbulence. The attempts to related turbulence to chaotic motion got strong impetus from the celebrated paper by Ruelle and Takens . Considerable success has been achieved mainly in the area: the onset of turbulence. For fully developed turbulence, many questions remain unanswered. The aim of this letter is to show that there are dynamical systems that are much simpler than the Navier-Stokes equations but that can still have turbulent states and for which many concepts developed in the theory of dynamical systems can be successfully applied. In this connection we advocate a broader use of the universal properties of a wide range of isotropic turbulence phenomena. Even for the case of fully developed turbulence, which contains an extreme range of relevant length scales, it is possible, by using the present model, to reproduce a surprising variety of relevant features, such as multifractal cascade, intermittency. This letter reverts to possible applications of the Navier-Stokes equations to studies of the nature of turbulence.

Turbulence is well deservedly often called " the last great unsolved problem of the classical physics". A comprehensive solution of the trublence problem from a physical point of view is still missing update. During the last few decades the theory of dynamical systems has experineced extremely rapid progress, due to the works of Lorenz, as well as Ruelle and Takens. For the understanding of turbulence, however, the success has been more limited. Turbulence, which implies spatial as well as temporal disorder, cannot be reduced to a low dimensional system, and thus a large part of the theory of dynamical systems, in particular regarding bifurcation structures and symbolic dynamics, is basically inapplicable. For fully developed turbulence, today's chaos theory is still not broadly enough developed.

The questions remaining open can be classified under two headings. On the one hand, we can enquire into the causes for the onset of turbulence. Why does a disordered, turbulent motion evolve from a laminar, smooth flow? Is the transition abrupt or gradual? The second group includes all those questions regarding ataractics of turbulent flows and manner in which the kinetic energy of the mean flow is distributed among the interlocking vortices and is finally converted into heat. Considerable success has been achieved mainly in the first area, the onset of turbulence. In the second area of interest, fully developed turbulence, many questions remain unanswered.

In the light of these remarks, the aim of this letter is to show that there are dynamical systems that are much simpler than the Navier-Stokes equations but that can still have turbulent states and for which many concepts developed in the theory of dynamical systems can be successfully applied. In this connection we advocate a broader use of the common properties of a wide range of natural phenomena. Even for the case of fully developed turbulence, which contains an extreme range of relevant length scales, it is possible, by using this Logistic map of isotropic turbulence, to reproduce a surprising variety of relevant features.

The Navier-Stokes equation probably contains all of turbulence. When put in an appropriate dimensionless form, the Navier-Stokes equations contain only one dimensionless parameter, the Reynolds number, defined by

$$R_e = \frac{UL}{n}$$

The Navier-Stokes equations are a dissipative dynamical system with many degrees of freedom, and the effective number of degrees of freedom increases strongly with increasing $R_e$. For low $R_e$, we observe and can compute smooth laminar solutions with little spatial structure. As $R_e$ increases, the generic sequence of events is a transition to a steady spatially structured flow, then to a spatially structured and time periodic flow, and then to a flow that is chaotic in time. This behavior is qualitatively similar for different flow geometries but is quantitatively far from universal.

Fluid turbulence provides a classic example where, as a parameter (the Reynolds number ) is tuned in a set of deterministic equations (the Navier-Stokes equations ), the motion can undergo an abrupt transition from some stable configuration ( for example, laminar flow ) into an apparently stochastic, chaotic regime. In this letter, it has observed that the Logistic map equation may be useful in this context. The main finds could be listed as follows:

**For isotropic turbulence, the corresponding Logistic map is**

$$x_{n+1} = ax_n(1 - x_n) \qquad \textcolor{blue}{\textbf{(1)}}$$

**where**

$$c = -2 + \frac{9}{2} \cdot \left(\frac{R}{R_c}\right)^{-\frac{3}{2}} - \frac{9}{4} \cdot \left(\frac{R}{R_c}\right)^{-3}$$

$$a = 1 + \sqrt{1 - 4c}$$

$x_n$ **is the average area of Taylor's micro-scale.** $R_c$ **is the critical value of Reynolds number for linear stability theory.**

The simple algebraic formula of Logistic map was introduced in 1845 by Verhulst to model the growth of the population of a biological species. In a fascinating and influential review article, Robert May (1976) emphasized that even this simple nonlinear maps could have very complicated dynamics. Figure 1-4 give some of details of the bifurcation of logistic map.

The logistic equation is one of the most often discussed prototypes of the complex behavior of deterministic systems. The best visible is a sequence of successive period doublings. It seems that periodic points first appear in order $1, 2, 4, 8, \ldots, 2^n, \ldots$ occur as $a$ increases. Specifically, let $\frac{R_m}{R_c}$ denote the value of bifurcation where a $2^n$ − cycle first appears. Then table.1 reveals that the critical bifurcation values of further period-doubling. Note that the successive bifurcations come faster and faster. Ultimately the bifurcation parameter converges to a limiting value $a_\infty$. For $a > a_\infty$, the orbit diagram reveals an unexpected mixture of order and chaos, with periodic windows interspersed between chaotic clouds of dots. This implies the transition from laminar to turbulence.

It should be note that what interesting us mostly is not only the route to turbulence, but also the universality of fully developed turbulence. During the last few years it has become clear that most fractals in nature are so-called multifractals. This means that the multifractal description gives a good summary of the observed statistical properties of the turbulent field. This phenomenological picture raises, however, as many physical questions as it answers. Starting from the Navier-Stokes equations, why should the statistics of the dissipation and the statistics of velocity field? In Fig.5 and Fig.6 we show $D_q$ and the $f(a)$ spectrum for isotropic turbulence at high Reynolds number based on the analysis of present theory.

In present framework, it is possible to analysis the link between anomalous scaling of the structure functions ( the nonlinear dependence on $p$ of the exponents $V_p$ ) and dynamical intermittency exhibited by the chaotic evolution. The relation between multifractality of energy dissipation in the 3D space and multifractality of the ergodic probability measure on the strange attractors in phase space is the major open problem of any theory of the scaling invariance in fully developed turbulence. The exponents $V_p$ can be expressed in terms of the generalized fractal dimensions $D_q$. Numerical simulations of the present model show that the exponents $V_p$ are not linear in $p$, and can be fitted by the She-Leveque formula given by phenomenology.

   In this letter, we have restricted the discussion to a particular turbulence problem, the putatively universal statistical properties of the small-scale velocity fluctuations in incompressible flows with high Reynolds number. We have found that the present theory might gives a compact description of a large amount of this turbulent problem. There is an adequate theoretical starting point in the Navier-Stokes equations. There is a useful picture that suggests at least approximate universality. Better mathematical understanding of the underlying Navier-Stokes equations will combine to increase our basic understanding.

Russian by V. I. Kisin. Mir Publishers.

ACKNOWLEDGMENTS. The work was supported by the National Natural Science Foundation of China (Grant Nos.10272018, 10572083).

PDF created with pdfFactory Pro trial version www.pdffactory.com

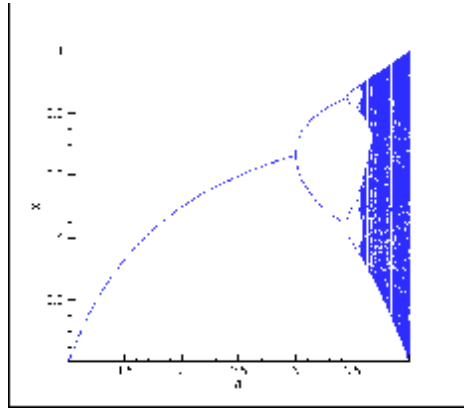

**Figure 1. The bifurcation diagram of logistic map**

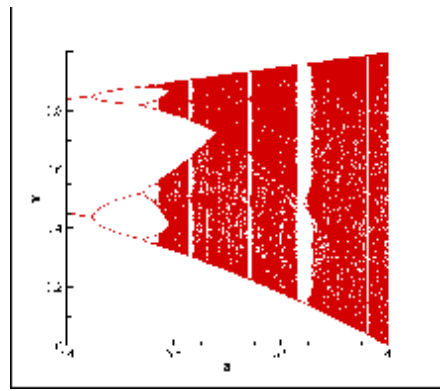

**Figure 2. Universality and self-smilarity of the bifurcation diagram of logistic map**

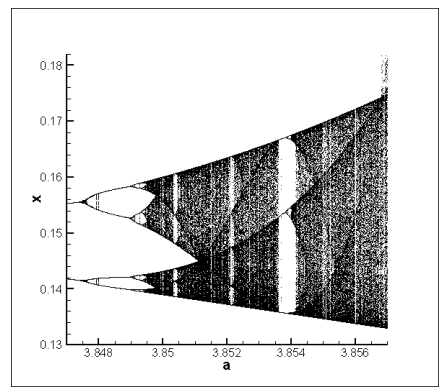

**Figure 3. Universality and self-smilarity of the bifurcation diagram of logistic map**

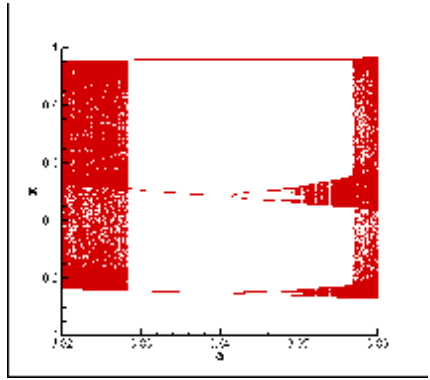

**Figure 4. Periodic windows : crises and self-similarity of logistic map**

**TABLE.1 The critical Reynolds number for period doubling bifurcation**

| $m$ | Period 2 is born | $\dfrac{R_m}{R_c}$ |
|---|---|---|
| 1 | $1 \to 2$ | 2.080083823 |
| 2 | $2 \to 2^2$ | 3.096733934 |
| 3 | $2^2 \to 2^3$ | 3.511518701 |
| 4 | $2^3 \to 2^4$ | 3.619876484 |
| 5 | $2^4 \to 2^5$ | 3.644190644 |
| 6 | $2^5 \to 2^6$ | 3.649451759 |
| ... | ... | ... |
| $\infty$ | $2^\infty \to$ chaos | 3.650890633 |

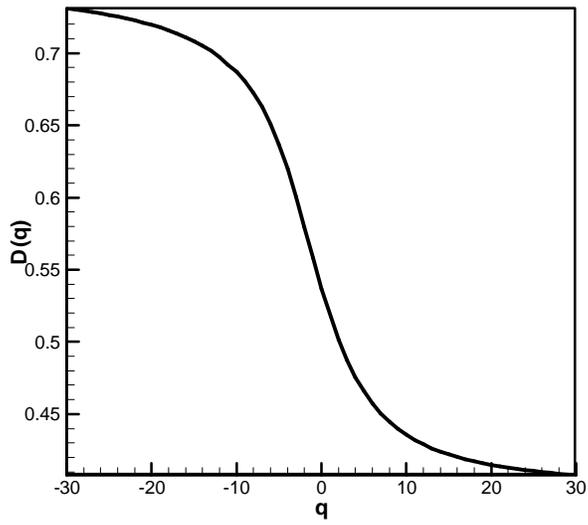

**Figure 5. Generalized dimensions** $D_q$ **for isotropic**

**Turbulence at high Reynolds number**

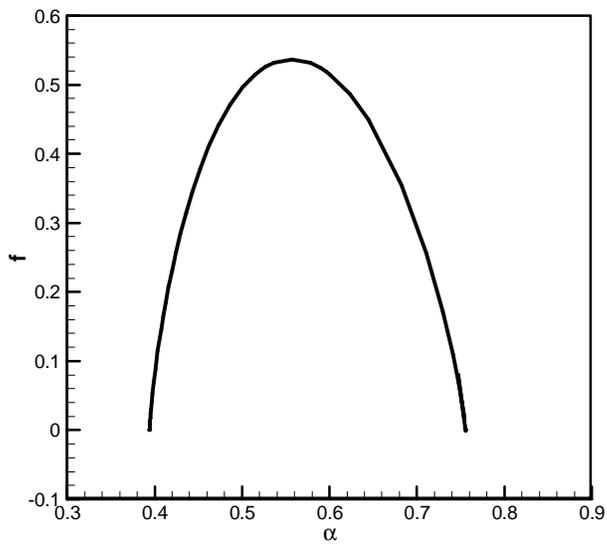

**Figure 6. Spectrum of scaling** $f(a)$ **for isotropic**

**Turbulence at high Reynolds number**

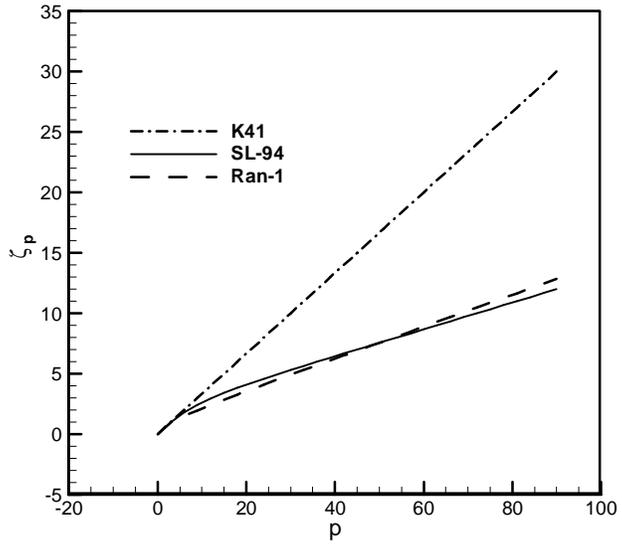

**Figure 7. The structure function exponents** $V_p$